\documentclass[12pt]{iopart}

\usepackage[english]{babel}
\expandafter\let\csname equation*\endcsname\relax
\expandafter\let\csname endequation*\endcsname\relax
\usepackage{amsmath}
\usepackage{graphicx}
\usepackage{xcolor}
\usepackage{float}
\usepackage[version=4]{mhchem}
\usepackage{siunitx}
\usepackage{tabularx}
\usepackage{dcolumn}
\usepackage{todonotes}
\usepackage{multirow} 
\usepackage{array} 

\begin{document}

\title[]{Comparative Performance of Fluorite-Structured Materials for Nanosupercapacitor Applications}

\author{Grégoire Magagnin}
\address{Ecole Centrale de Lyon, INSA Lyon, CNRS, Universite Claude Bernard Lyon 1, CPE Lyon, INL, UMR5270, 69130 Ecully, France}
\author{Jordan Bouaziz}
\address{Ecole Centrale de Lyon, INSA Lyon, CNRS, Universite Claude Bernard Lyon 1, CPE Lyon, INL, UMR5270, 69130 Ecully, France}
\ead{jordan.bouaziz@ec-lyon.fr}
\author{Martine Le Berre}
\address{INSA Lyon, Ecole Centrale de Lyon, CNRS, Universite Claude Bernard Lyon 1, CPE Lyon, INL, UMR5270, 69621 Villeurbanne, France}
\author{Sara Gonzalez}
\address{CNRS, INSA Lyon, Ecole Centrale de Lyon, Universite Claude Bernard Lyon 1, CPE Lyon, INL, UMR5270, INSA, Bât. Irène Joliot-Curie, 3 rue Enrico Fermi, F-69621 Villeurbanne Cedex, France}
\author{Damien Deleruyelle}
\address{INSA Lyon, Ecole Centrale de Lyon, CNRS, Universite Claude Bernard Lyon 1, CPE Lyon, INL, UMR5270, 69621 Villeurbanne, France}
\author{Bertrand Vilquin}
\address{Ecole Centrale de Lyon, INSA Lyon, CNRS, Universite Claude Bernard Lyon 1, CPE Lyon, INL, UMR5270, 69130 Ecully, France}

\begin{abstract}
    Over the last fifteen years, ferroelectric and antiferroelectric ultra thin films based on fluorite-structured materials have drawn significant attention for a wide variety of applications requiring high integration density. Antiferroelectric \ce{ZrO_2}, in particular, holds significant promise for nanosupercapacitors, owing to its potential for high energy storage density (ESD) and high efficiency ($\eta$). This work assesses the potential of high-performance \ce{Hf_{1-x}Zr_{x}O_2} thin films encapsulated by TiN electrodes that show linear dielectric (LD), ferroelectric (FE), and antiferroelectric (AFE) behavior. Oxides on silicon are grown by magnetron sputtering and plasma-enhanced atomic layer deposition. The electric behavior of the selected samples is enlightened by the corresponding crystalline phases observed in X-ray diffraction. A wake-up (WU) effect is observed for AFE \ce{ZrO_2}, a phenomenon that has been barely reported for pure zirconium oxide and AFE materials in general. This WU effect is correlated to the disappearance of the pinched hysteresis loop (related to the superposition of a double peak in I-V curves) commonly observed for \ce{Zr}-doped \ce{HfO_2} thin films after WU. ESD and $\eta$ are compared for FE, AFE, and LD samples at the same electrical field (\SI{3.5}{MV/cm}). As expected, ESD is higher for the FE sample (\SI{95}{J/cm^3}), but $\eta$ is ridiculously small ($\approx$ \SI{55}{\percent}), because of the opening of the FE hysteresis curve inducing high loss. Conversely, LD samples exhibit the highest efficiency (nearly \SI{100}{\percent}), at the expense of a lower ESD. AFE \ce{ZrO_2} thin film strikes a balance between FE and LD behavior, showing reduced losses compared to the FE sample but an ESD as high as \SI{52}{J/cm^3} at \SI{3.5}{MV/cm}. This value can be further increased up to \SI{84}{J/cm^3} at a higher electrical field (\SI{4.0}{MV/cm}), with an $\eta$ of \SI{75}{\percent}, among the highest values reported for fluorite-structured materials, offering promising perspectives for future optimization.
\end{abstract}

\section{Introduction}

\indent 
    Immediate remedies are essential to address the challenges posed by the exponential increase of energy consumption. Specifically, pivotal technologies related to the fourth industrial revolution—such as the Internet of Things and Big Data—are witnessing an exponential surge in energy consumption linked to the storage, processing, and transmission of digital information \cite{IEA2017digitalisation}. Harnessing the potential of ferroelectric (FE) and antiferroelectric (AFE) materials compatible with complementary metal–oxide–semiconductor (CMOS) technology is a compelling strategy for the creation of energy-efficient electronic devices more specifically for energy conversion applications.

\indent 
    The term "Fluorite structure" denotes a prevalent pattern observed in compounds represented by the formula \ce{MX_2}. In this arrangement, the X ions are situated in the eight tetrahedral interstitial sites, while the M ions occupy the regular sites within a face-centered cubic structure. This structural configuration is commonly observed in various compounds, notably the mineral fluorite (\ce{CaF_2}) which gave its name to the structure. Typical fluorite-structured ferroelectrics and antiferroelectrics are respectively doped hafnium oxide (\ce{HfO_2}) and zirconium oxide (\ce{ZrO_2}). \ce{HfO_2} has been introduced since 2007 by Intel as a high-k (high dielectric constant) in the gate stack of MOS transistors (metal oxide semiconductor) \cite{mistry200745nm} while \ce{ZrO_2} is also widely used to fabricate DRAM cells \cite{james2013recent}. In 2011, a ferroelectric phase in Si-doped \ce{HfO_2} was first reported \cite{boscke2011ferroelectricity,boescke2010integrated}, followed by the discovery of ferroelectricity in a solid solution of \ce{Hf_{0.5}Zr_{0.5}O_2} (HZO) the same year \cite{muller2011ferroelectric}, paving the way for the re-introduction of ferroelectric materials in the existing CMOS technology.
    
\indent 
    At atmosphere pressure, bulk \ce{HfO_2} and \ce{ZrO_2} have centrosymmetric non-polar crystal structures \cite{wang1992hafnia}. However, under certain conditions of TiN encapsulation, doping and/or film stress, it is possible to stabilize a meta-stable orthorhombic phase, which gives rise to ferroelectricity (FE) in \ce{HfO_2} and in \ce{HfO_2}-\ce{ZrO_2} thin film solid solutions \cite{park2015ferroelectricity}. Contrary to antiferroelectricity originating from anti-parallel dipole moments arrangements, the origin of the AFE properties in \ce{ZrO_2} is attributed to the electric field-induced structural phase transitions between the non-polar tetragonal phase and the polar orthorhombic phase, ensuing in a double hysteresis polarization loop (PE loop) \cite{lomenzo2022harnessing, lomenzo2023discovery}.
    
\indent 
    FE and AFE nanosupercapacitors can be used for solid-state electrostatic energy storage \cite{hao2013review} where high energy storage performances has been reported in ferroelectric \ce{HfO_2} or \ce{ZrO_2} films \cite{park2014thin,hoffmann2019negative}. The field-induced transitions observed in AFE \ce{ZrO_2} hold promise for energy storage applications. However, several physical parameters from the fluorite films can limit the energy storage performances. Ferroelectric thin films exhibit a “wake-up” (WU) effect, which corresponds to an increase of the remnant polarization with cycling. This effect depends on the amplitude and frequency of the cyclic applied voltage stress \cite{zhou2013wake}. Another limiting factor is the film thickness scaling. Fluorite films have limited energy storage scaling properties due to the increase of the monoclinic phase proportion at large thicknesses \cite{park2013evolution}.

\indent
     Ferroelectrics (FE) excel in achieving high polarization, leading to high energy storage density (ESD). However, they have rather low efficiency, while linear dielectrics (LD) demonstrate remarkable efficiency, but low polarization \cite{QI2022local}. In the context of prior research, it becomes evident that antiferroelectrics (AFE) offer an optimal balance, offering superior characteristics by combining elevated ESD and higher efficiency. Surprisingly, the exploration of these distinct attributes within the same material, thickness, and capacitor structures has been notably limited. In this context, this work aims at addressing this gap, employing \ce{ZrO_2} and HZO as prototype materials for a comprehensive investigation. This study proposes a performance assessment and comparison between AFE, FE and LD fluorites for nanosupercapacitor applications. Fluorite thin films were grown with the same chemical composition but with different deposition techniques and parameters, leading to different nanosupercapacitor electrical properties, from LD to FE and AFE. AFE fluorite thin films exhibit beyond state-of-the-art energy storage capabilities.

\section{\label{sec:Method}Materials and methods}

\indent 
    Capacitors were fabricated on p-doped \ce{Si} (001) substrates and follow the stack Pt/TiN/oxide/TiN/Si. Details of the deposition processes for oxides are summarized in table \ref{tab:growth_cond}. Sputtering is performed via AC450 magnetron sputtering chamber from Alliance Concept, while Plasma-Enhanced Atomic Layer Deposition (PEALD) is carried by a Fiji F200 apparatus from Ultratech. The first step of \ce{ZrO_2} PEALD deposition consists in the application of an \ce{O_2} plasma on the TiN bottom electrode at \SI{300}{\watt}, before opening the valve of the \ce{TDMA}-\ce{Zr} precursor. Alternation of a complex sequence, notably including the \ce{TDMA}-\ce{Zr} valve opening and the dioxygen plasma is then performed to grow the AFE \ce{ZrO_2} thin film. Sputtering is performed for TiN and Pt using a metallic target of \ce{Ti} and \ce{Pt}. \ce{Pt}/\ce{TiN} top electrodes are obtained after a photolithography and lift-off process. Rapid thermal annealing (RTA) was then performed for all samples. Samples where then investigated by means of physical and electrical characterization.

\begin{table}[]
\resizebox{\textwidth}{!}{%
\begin{tabular}{c||ccc||c}
Oxide  & \ce{HZO}(LD)            & \ce{HZO}(FE)               & \ce{ZrO_2}(LD)          & \ce{ZrO_2}(AFE)                   \\ \hline \hline
Deposition Technic            & Reactive sputtering & Non-Reactive Sputtering & Reactive Sputtering & PEALD                        \\
Working Pressure (mbar)              & \SI{5e-3}{}                & \SI{5e-2}{}                     & \SI{5e-3}{}                 & \SI{5e-1}{}                          \\
Target or Precursors          & \ce{Hf}/\ce{Zr}               & \ce{HfO_2}/\ce{ZrO_2}               & \ce{Zr}                  & \ce{TDMA}-\ce{Zr}                      \\
Deposition Temperature        & \multicolumn{3}{c||}{Room Temperature}    & \SI{200}{\celsius}                        \\
RTA                           & \multicolumn{3}{c||}{ \SI{450}{\celsius} - \SI{30}{\second} - \ce{N_2} atmosphere}                    & \SI{600}{\celsius}  - \SI{30}{\second} - \ce{N_2} atmosphere \\
Thickness   (nm)                  & 9.1               & 13.0                  & 8.5                & 10.3                      
\end{tabular}%
}
\caption{Growth conditions of the oxide thin films. Electrical properties of the films are shown in figure \ref{fig:P-E_and_I-V} and given in the brackets.}
\label{tab:growth_cond}
\end{table}

\indent 
    Glancing Incidence X-ray Diffraction (GIXRD) was performed on a Smartlab Rigaku diffractometer using a \SI{9}{kW} copper rotating anode, a parabolic multi layer mirror
    for parallel beam setting, Ni filter for \ce{Cu}$K_\alpha$ radiation selection, 0.114°
    aperture parallel slit analyser, and a 0D scintillating counter. The thickness of all thin films was measured by X-Ray Reflectivity (XRR) with the same instrument.

\indent
    Electrical characterization was carried out on \SI{50}{\micro \meter} and \SI{20}{\micro \meter} diameter capacitors using a probe station in a Faraday cage and a setup composed of a Keithley 4200SCS equipped with PMU. Endurance tests were performed using a custom program interfaced with the Keithley. The cycling sequence consists in bipolar voltage square pulses (commonly called set/reset sequence) until breakdown at \SI{3.5}{\volt} to \SI{4.5}{\volt} depending on the film thickness with \SI{20}{\micro \second} pulse duration. Polarization as a function of electric field (P-E) curves are established from the measurement sequence consisting 3 triangle pulses (DHM : Dynamic Hysteresis Measurement) with a \SI{60}{\micro \second} rise time. The first voltage pulse poles the polarization in a given pre-set direction, while the two other pulses measures the current response. The pulse amplitude is set to obtain a \SI{3.5}{MV/cm} electrical field for each samples, and for all figures, except figure \ref{fig:ESD_vs_Eff}, where a maximum electrical field of \SI{4.0}{MV/cm} is applied in order to maximize the ESD of the AFE \ce{ZrO_2}.

\section{Results and Discussion}

\begin{figure}[H]
    \centering
    \includegraphics[width=1\textwidth]{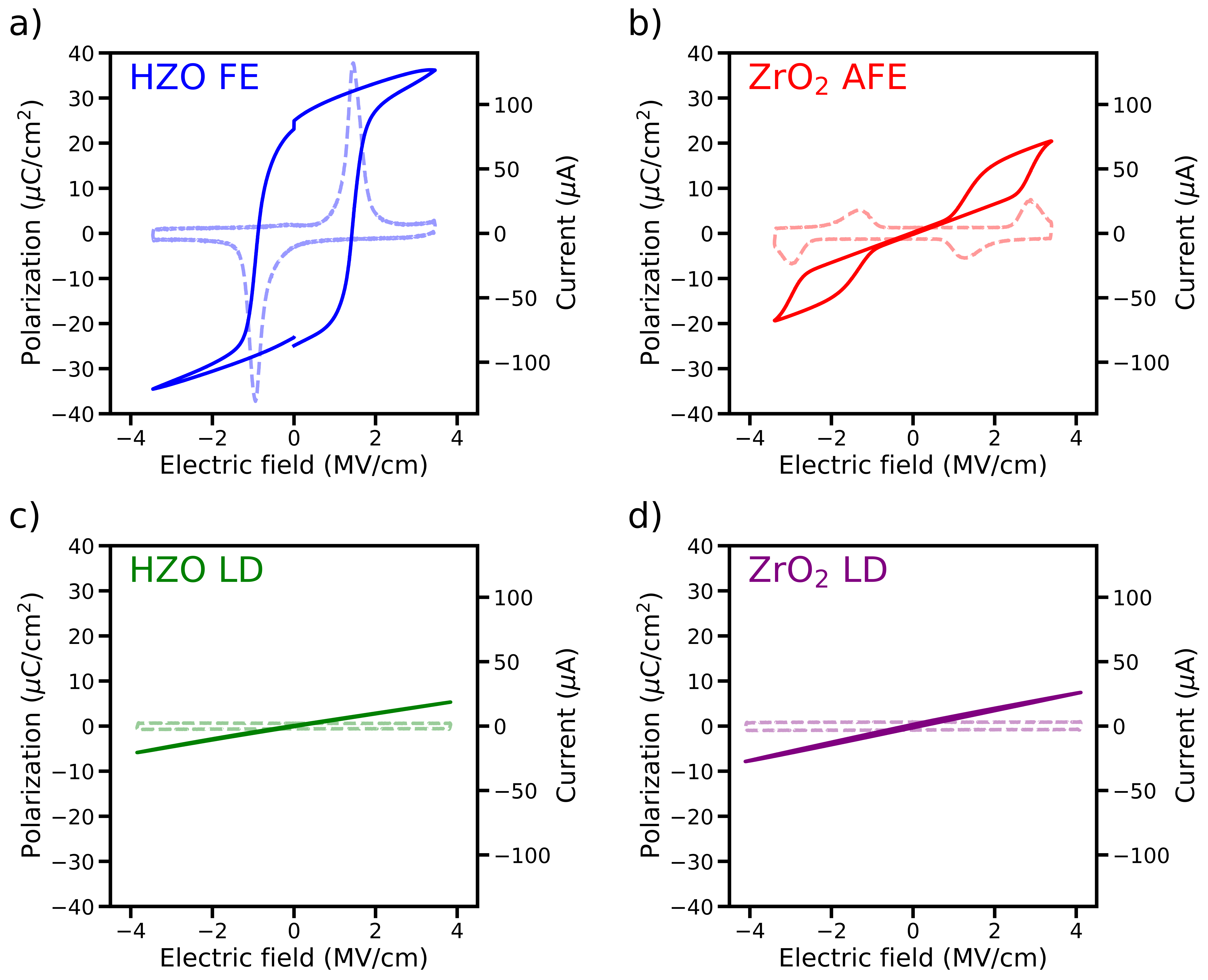}
    \caption{\label{fig:P-E_and_I-V}Polarization versus Electrical field (straight lines) and Current versus Voltage (dashed lines) after $10^3$ cycles for a) a ferroelectric $\mathrm{Hf}_{0.5}\mathrm{Zr}_{0.5}\mathrm{O}_2$ (FE HZO), b) an anti-ferroelectric $\mathrm{ZrO}_2$ (AFE \ce{ZrO_2}), c) a linear dielectric $\mathrm{Hf}_{0.5}\mathrm{Zr}_{0.5}\mathrm{O}_2$ (LD HZO) and d) a linear dielectric $\mathrm{ZrO}_2$ (LD \ce{ZrO_2}).}
\end{figure}

\indent 
    The field induced electrical properties of the fluorite thin film capacitors grown by sputtering and ALD were first examined. Fig \ref{fig:P-E_and_I-V} shows the electrical characteristics of the capacitors comparing FE and AFE samples with LD ones of the same chemical composition, at the same applied electric field of 3 MV/cm. On Figure \ref{fig:P-E_and_I-V}, FE, AFE and LD properties of \ce{HZO} and \ce{ZrO_2} are shown after $10^3$ cycles. Polarization versus electrical field (solid curves) and current versus voltage (dashed curves) are systematically shown for each studied samples. 

\indent     
    For LD samples on Figure \ref{fig:P-E_and_I-V}(c) and (d), as the dielectric capacitor is charging or discharging, the current is different from zero, leading to a non-zero electric displacement field for non-zero electric field. Therefore, a linear relationship between polarization and the applied electric field is expected. Due to leakage currents, the P-E loop is not totally closed and will lead to an efficiency of the energy storage close but less than \SI{100}{\percent} (the theoretical value for a perfect LD).
    
\indent    
    For the FE HZO on Figure \ref{fig:P-E_and_I-V}(a), after $10^3$ cycles, two peaks can be observed: one at positive voltages and one at negative voltages. These peaks corresponds to polarization switching peaks, due to the FE nature of the film, attributed to the displacement of oxygen ions in HZO \cite{materlik2015origin}. The remnant polarization of the FE HZO is \SI{23}{\micro \coulomb / \centi \meter ^2} on the positive side and \SI{24}{\micro \coulomb / \centi \meter ^2} on the negative side. The small asymmetry is attributed to the possible oxidation state of the top \cite{segantini2023interplay} or bottom \cite{bouaziz2019dramatic} TiN electrode. 
    
\indent    
    Finally, for the AFE \ce{ZrO_2} sample, a state-of-the-art curve is observed after $10^3$ cycles. A threshold field of about \SI{1.0}{MV/cm} can be seen between an LD to FE behavior, corresponding to the field induced phase transition assumed for \ce{ZrO_2} \cite{hoffmann2022antiferroelectric, lomenzo2023discovery}. A very sharp linear opening is present below \SI{1.0}{MV/cm} followed by a narrow hysteresis loop above, with a saturation polarisation $P_s$ as high as \SI{20.5}{\micro \coulomb / \centi \meter ^2}. 
    
\begin{figure}[H]
    \centering
    \includegraphics[width=1\textwidth]{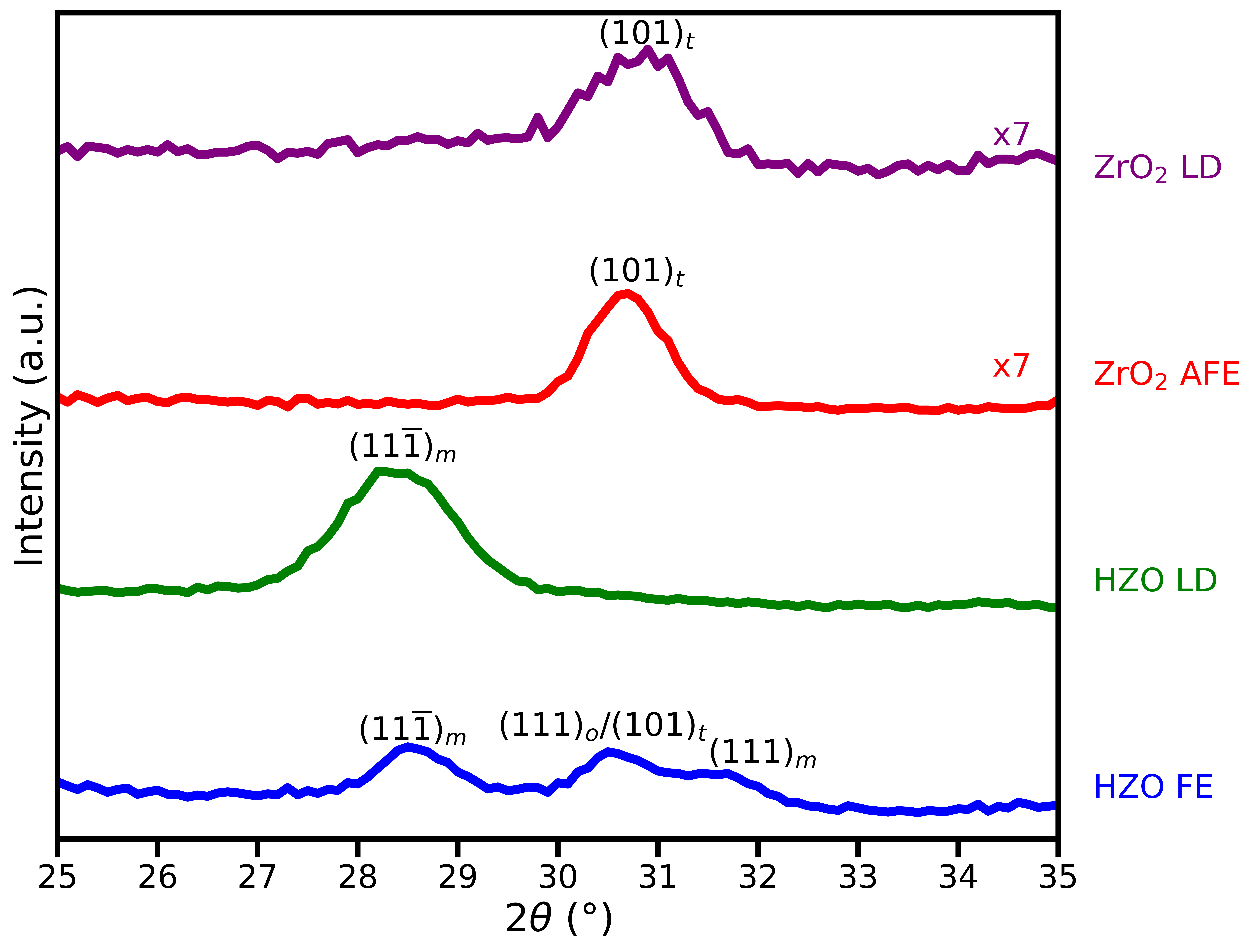}
    \caption{\label{fig:XRD} GIXRD scans of FE \ce{HZO} (red curve), AFE \ce{ZrO2} (blue curve), LD \ce{ZrO2} (green curve), LD \ce{HZO} (purple curve).}
\end{figure}

\indent
    Structural characterization measurements are then conducted on each sample to elucidate the origin of the functional properties allowing for energy storage. Figure \ref{fig:XRD} shows the XRD scans for all FE, AFE and LD samples. The FE \ce{HZO} sample exhibits a distinct orthorhombic/tetragonal (o/t) peak with (111) orientation for the o-phase and (101) orientation for the t-phase around a $2\theta$\ value of \SI{30.5}{\degree}. The non-centrosymmetric o-phase is typically considered the phase responsible for ferroelectricity. However, it also shows the presence of the monoclinic (m-) phase, which is centrosymmetric. The mixture of o/t phases is attributed to \ce{HZO}, while only the (101)-oriented t-phase is attributed to \ce{ZrO_2} for the peak around \SI{30.5}{\degree}, considering current literature explanations \cite{park2018review,hoffmann2022antiferroelectric}. 

\indent    
    Properties of HZO thin films of (approximately \SI{10}{\nano \meter} thick), synthesized via reactive magnetron sputtering from a \ce{Hf}/\ce{Zr} metallic target \cite{bouaziz2019characterization,bouaziz2019dramatic} on a TiN layer, were explored. GIXRD measurements revealed that, depending on the deposition working pressure in the chamber – low pressure (LP) (\SI{5e-3}{mbar}) or high pressure (HP) (\SI{5e-2}{mbar}) – the thin films were either monoclinic or amorphous after deposition. The \ce{ZrO_2} films grown by ALD \cite{muller2011ferroelectric} are amorphous after deposition. Amorphous samples exhibit the o/t peak after Rapid Thermal Annealing (RTA), while monoclinic samples remain in their monoclinic structure after RTA. 
    
\indent   
    For the \ce{HfO_2}/\ce{ZrO_2} ceramic target and the \ce{Zr} metallic target used in this study (respectively non-reactive and reactive magnetron sputtering), regardless the pressure, the \ce{HZO} and \ce{ZrO_2} thin films are amorphous after deposition. After RTA, films obtained at both pressures exhibit the o/t peak, but only the HZO HP samples also show some monoclinic peaks, whereas LP samples only show the o/t peak (not shown in this paper). It was demonstrated that \ce{HZO} LP samples are only tetragonal \cite{cervasio2024quantification} while HP samples show a mixture of tetragonal and orthorhombic phase (in addition to the monoclinic phase for HP samples). 
    
\indent      
    While all samples present the o/t peak in their GIXRD scans, their electrical behaviors are vastly different. Our observations tend to show that observing a peak around \SI{30.5}{\degree} is a necessary condition to have FE or AFE properties, but it is not a sufficient condition. Structural measurements alone are generally insufficient to clearly identify the electrical nature of the thin films. Therefore, electrical characterizations in Figure \ref{fig:P-E_and_I-V} are needed to conclude about the electric nature of the \ce{HZO} and \ce{ZrO_2} capacitors. 

\indent 
    Endurance tests were also performed and a wake-up effect was observed. Historically, the first observation of a wake-up effect dates back to Sim et al. \cite{sim2008ferroelectric}. However, in Sim et al. article, the increase in $P_r$ seems to be attributable to the increase in leakage current (fatigue phenomena) at increasing cycling counts. There is no evidence that this effect could be similar to that observed for \ce{HfO_2}. In 2011, Wu et al. \cite{wu2011compositionally} continued the work of Sim et al. They coined, for the first time, the increase in Pr with the number of cycles as the "wake-up" effect. However, here again, this so-called "wake-up" effect can be attributed to the increase in leaks and phenomena of modification of space charge. In the same article, it is also noted that if the frequency decreases, $P_r$ increases. In 2012, finally, Mueller et al. \cite{mueller2012reliability} released the first article discussing the "wake-up" effect on \ce{Si}:\ce{HfO_2}. They named this effect the wake-up effect in reference to the endurance test procedure carried out by Wu et al. and then spoke of a "wake-up procedure" and not a "wake-up effect." In 2013, Zhou et al. \cite{zhou2013wake} were the first to truly study the wake-up (WU) effect and to name it as such. This time, the argument of increased leaks is no longer mentioned, although this is not conclusively proven in this article. Thanks to the electrical characterization PUND technique, it was already observed in previous works that the WU effect does not result from an increase in leakage currents \cite{bouaziz2019huge}. Moreover, Zhou et al. observe that as the measurement frequency increases, Pr decreases, while it would increase at increasing pulse voltage amplitudes. It will be shown later that a WU effect is also present for AFE \ce{ZrO_2}, a detailed observation of the wake up and endurance properties of the analyzed films is present in the supplemental materials. This WU effect, although almost identical from the perspective of current shifts along cycling in FE HZO and AFE \ce{ZrO_2}, cannot be defined as an increase of $P_r$. 

\begin{figure}[H]
    \centering
    \includegraphics[width=1\textwidth]{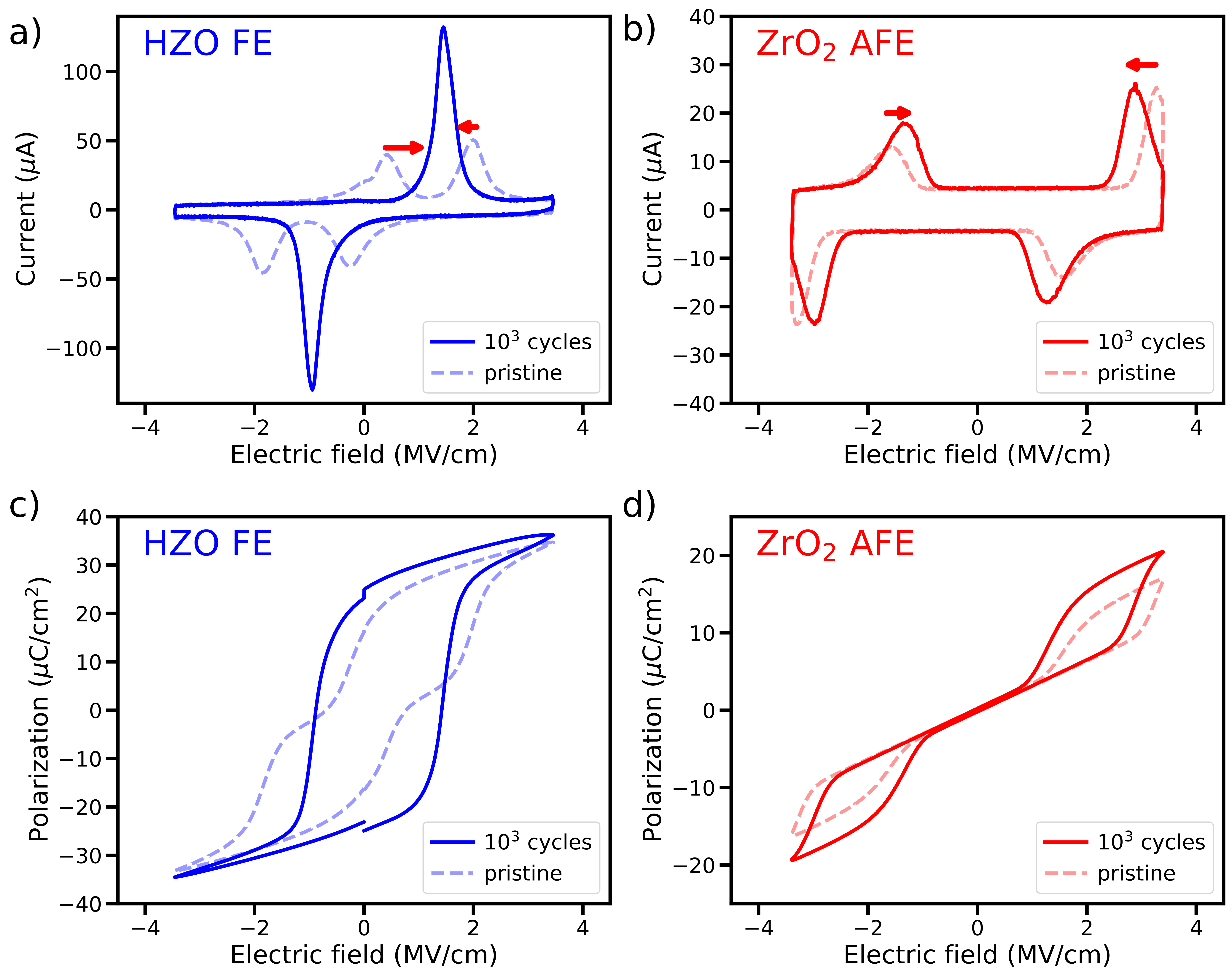}
    \caption{\label{fig:WU}Current versus electric field measurements for pristine samples (dashed lines) and $10^3$ cycles (straight lines) of (a) FE \ce{HZO} (b) AFE \ce{ZrO_2} and their corresponding polarization versus electric field measurements of (c) FE \ce{HZO} (d) AFE \ce{ZrO_2}. The observed change in current between pristine and $10^3$ cycled samples correspond to the wake-up effect (WU), resulting in the switching current peaks shift over the voltage axis highlighted by the red arrows.}
\end{figure}
    
\indent 
    Figure \ref{fig:WU}a) and b) depict current versus voltage (I-V) curves for FE \ce{HZO} and AFE \ce{ZrO_2}, respectively, whereas Figure \ref{fig:WU}c) and d) illustrate P-E loops for the same samples. Dashed lines represent the behavior of the samples in their pristine state, while solid lines indicate their behavior after $10^3$ endurance cycles. At the pristine stage, FE \ce{HZO} exhibits 4 switching current peaks : one pair at positive voltages and another one at negative voltages (dashed lines Figure \ref{fig:WU}(a)). Along cycling, each pair would progressively merge (solid lines Figure \ref{fig:WU}(a)), leading to the hysteresis loop on Figure \ref{fig:P-E_and_I-V}c) (solid lines). For simplicity, observed peaks at the pristine stage for the FE \ce{HZO} will be referred as "double peak" or "double peak phenomenon" from now on. For positive voltages, the left peak shifts towards higher voltage values, whereas the right peak shifts towards lower voltage values, as shown by the red arrows. In terms of FE domains, this implies that certain FE domains are undergoing switching at lower voltage levels, while others are switching at higher values. With an increasing number of cycles, low-voltage switching domains transition to higher voltage values, while high-voltage switching domains transition to lower values, eventually resulting in the merging of the two peaks. In FE HZO thin films grown by sputtering, this effect has been already well described \cite{bouaziz2019huge,manchon2022insertion}. 

\begin{figure}[H]
    \centering
    \includegraphics[width=1\textwidth]{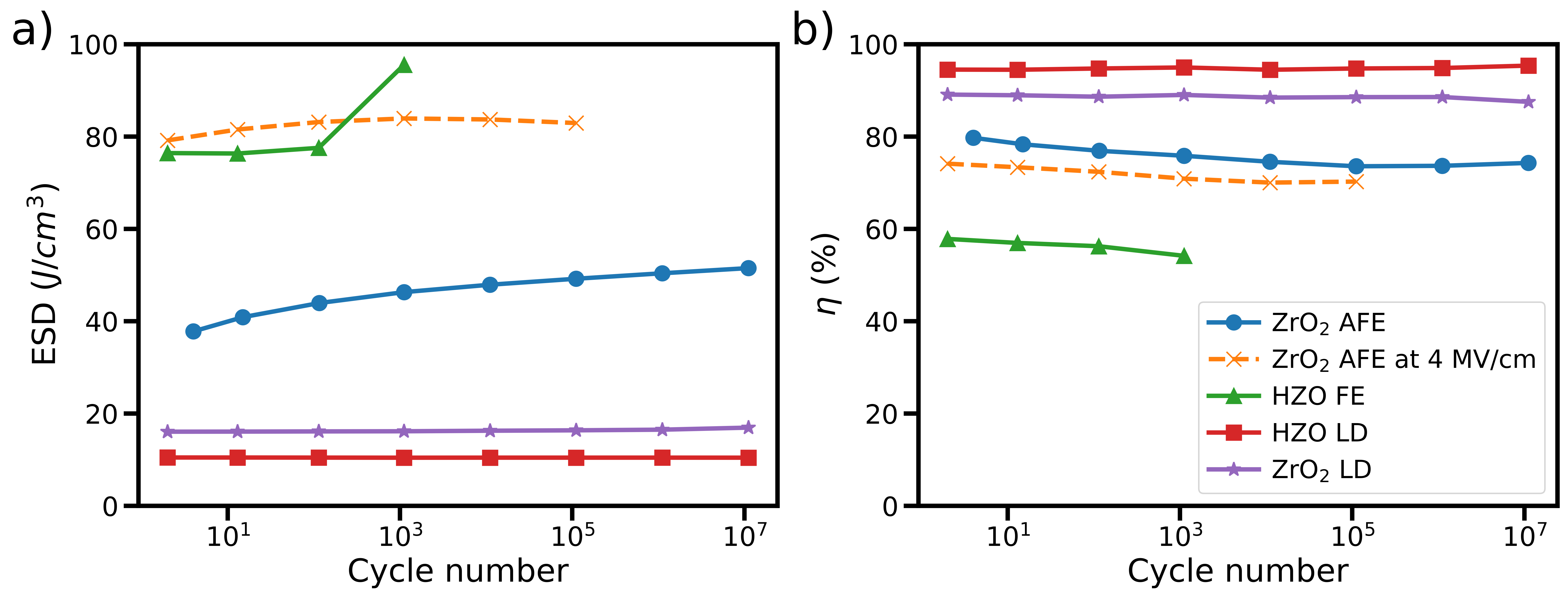}
    \caption{\label{fig:ESD_and_Eff_vs_cycles}(a) Energy density storage versus cycles and (b) efficiency ($\eta$) versus cycles until breakdown for FE \ce{HZO}, AFE \ce{ZrO_2}, LD \ce{HZO} and LD \ce{ZrO_2} at an applied electric field of 3.5 MV/cm. For comparison, AFE \ce{ZrO_2} at 4 MV/cm is also shown.}
\end{figure}

\indent
    Although WU effect is rarely mentioned for AFE, we observed similar current peak displacement for \ce{ZrO_2} than for FE HZO as highlighted by the red arrows on figure \ref{fig:WU}(a) and (b). In contrast to FE \ce{HZO}, the pristine values for AFE \ce{ZrO_2} exhibit different signs. It has to be mentioned that $P_r$ doesn't apply for AFE, since around zero volt AFE are showing the same behavior as LD. Nevertheless, on figure \ref{fig:WU}(d), the two hysteresis loops of the AFE PE curves have smaller coercive fields and a higher $P_s$ between pristine and woken states, leading to an increase in the ESD on Figure \ref{fig:ESD_and_Eff_vs_cycles}(a) as $P_s$ increases and loss decreases. 

\indent 
    This intriguing similarity between the double peak phenomenon in FE and the AFE behavior has already been discussed \cite{lomenzo2023discovery}. The non-uniform distribution of the internal electric field is likely attributed to unevenly distributed charged defects, such as oxygen vacancies, particularly near the electrodes. This asymmetry in oxygen vacancy concentration, often induced by the reduction of the doped \ce{HfO_2} layer by metal nitride electrodes, is a potential source of the internal field in the pristine material. The non-uniform distribution of oxygen vacancies near the electrodes may create an asymmetric internal field. In the process of electric field cycling, oxygen vacancies might diffuse into the bulk regions of fluorite-based films, triggering the wake-up process and resulting in the merging of switching current peaks in the case of FE \ce{HZO}. Subsequent investigations have reported a redistribution of charges associated with oxygen vacancies \cite{pevsic2016physical, Hamouda2022oxygen}. Another plausible mechanism for the wake-up effect involves field-cycling-induced phase transitions \cite{kim2016study}. Lomenzo et al. \cite{lomenzo2022harnessing} initially proposed that the transition from the tetragonal (t-) to the ferroelectric orthorhombic phase (o-phase) underlies the wake-up effect. They observed a decrease in dielectric permittivity and an increase in $P_r$ with an increasing number of electric field cycles, possibly indicating a phase transition from a non-ferroelectric phase with higher permittivity to a ferroelectric phase with lower permittivity. Additionally, Grimley al. employed scanning transmission electron microscopy (STEM) and impedance spectroscopy to observe a phase transition from monoclinic (m-) to o-phase \cite{grimley2016structural}. In essence, phase change and the redistribution of defects can induce the pinning of domains, leading to WU in both FE and AFE \cite{fengler2017domain,fengler2018relationship}. 
    
\indent
    Despite the fact that some authors have considered that the double peak is not similar to AFE phenomena \cite{lomenzo2023discovery}, the double peak on the positive voltage side can be considered as resulting from interactions between some negatively charged regions and positively charged regions or screened regions that define the switching current at the pristine stage. And after cycling, domains tend to homogenize, similarly to the observed phenomena for both FE \ce{HZO} and AFE \ce{ZrO_2}. Further investigations on the microstructure and oxygen vacancies re-organization would help to clearly determine if both phenomena have the same origin or not. 

\indent
    FE and AFE can reach high polarization values for low applied electric fields compared to dielectrics. This material functional property can be tailored for embedded energy storage capacitors of nanometer size that can reach high current density storage with low losses for low applied electric field. The electrical storage can be assessed on the thin film capacitor by calculating the energy storage density. By definition, the total energy $W$ stored in a capacitor (expressed in joules) is the total work done in establishing the electric field from an uncharged state \cite{Purcell_Morin_2013} :

\begin{equation}
    W = \int_0^Q V(q) dq \label{eq:ES}
\end{equation}
    
    By considering geometry factors in our Metal/Insulator/Metal (MIM) capacitors: the thickness of the insulator and the surface of the electrodes, it leads to the following expression of that the energy-storage density (ESD):
    
\begin{equation}
    W_{ESD} = \int_0^{E_{max}} PdE \text{ (upon discharging)}
\end{equation}

    where it is considered that ESD is equal to $W_{ESD}$. As previously said, the definition for the total energy stored in equation \ref{eq:ES} is calculated upon charging, starting from an uncharged state, while ESD is calculated upon discharging, because the definition considers a perfect linear dielectric and therefore it does not take the losses (due to leakage current in the case of a LD) into account. Then, the loss can be calculated as :
    
\begin{equation}
    W_{loss} = \int_0^{E_{max}} PdE \text{ (upon charging)} - W_{ESD}
\end{equation}

As a consequence the efficiency (in percentage) of the charge/discharge is given by: 

\begin{equation}
    \eta = \frac{W_{ESD}}{W_{ESD} + W_{loss}} \times 100
\end{equation}

    These calculations for ESD, loss and $\eta$ are now standard performance indicator for fluorite-based capacitors. \cite{hao2013review,Yi2021ultra}.

\indent
    Figure \ref{fig:ESD_and_Eff_vs_cycles} shows the ESD and $\eta$ as a function of the number of cycles for an applied field of 3.5 MV/cm on the four analyzed samples and also for \ce{ZrO_2} cycled at 4 MV/cm. Contrary to the \ce{HZO} FE layer, the breakdown field of the \ce{ZrO_2} AFE layer is higher allowing to display the film properties at 4 MV/cm. As expected, on Figure \ref{fig:ESD_and_Eff_vs_cycles}(a) one can observe the very high energy density of AFE \ce{ZrO_2} compared to the other samples. For LD \ce{HZO} and \ce{ZrO_2}, ESD is very low due to the low values of polarization when applying an electric field. But the samples have better endurance properties as they are not experiencing breakdown at $10^7$ cycles, contrary to FE \ce{HZO} and AFE \ce{ZrO_2}. For FE \ce{HZO}, at the early stage of cycling, the capacitor is very similar to AFE \ce{ZrO_2}, because of the double peak phenomena. But as the number of cycles increases, switching peaks start to merge and ESD is therefore increasing, leading to a higher ESD for FE \ce{HZO} than the one of AFE \ce{ZrO_2} at $10^3$ cycles. 
    
\indent
    On Figure \ref{fig:ESD_and_Eff_vs_cycles}(b) as one could expect the most efficient samples are the LD ones. One can observe that $\eta$ is actually not totally equal to \SI{100}{\percent}, because of leakage currents and have the lowest ESD of all samples. 
    As the hysteresis of FE \ce{HZO} is wide opened, it has the highest losses, hence it shows the lowest $\eta$ of the four samples of $\approx$ \SI{55}{\percent}. At the same time, the high $P_s$ of the FE sample make it reach the highest ESD value at \SI{95}{J/cm^3}.
    Finally, the reason why AFE are considered as a better option for supercapacitors compared to simple FE and LD inorganic electrostatic capacitors, is because of their efficiency falling in between FE and LD, with an $\eta$ of \SI{75}{\percent} while their ESD is almost as high as the one of FE samples, reaching \SI{52}{J/cm^3} at \SI{3.5}{MV/cm}. This value can be further increased up to \SI{84}{J/cm^3} at a higher electrical field (\SI{4.0}{MV/cm}). 

\begin{figure}[H]
    \centering
    \includegraphics[width=1\textwidth]{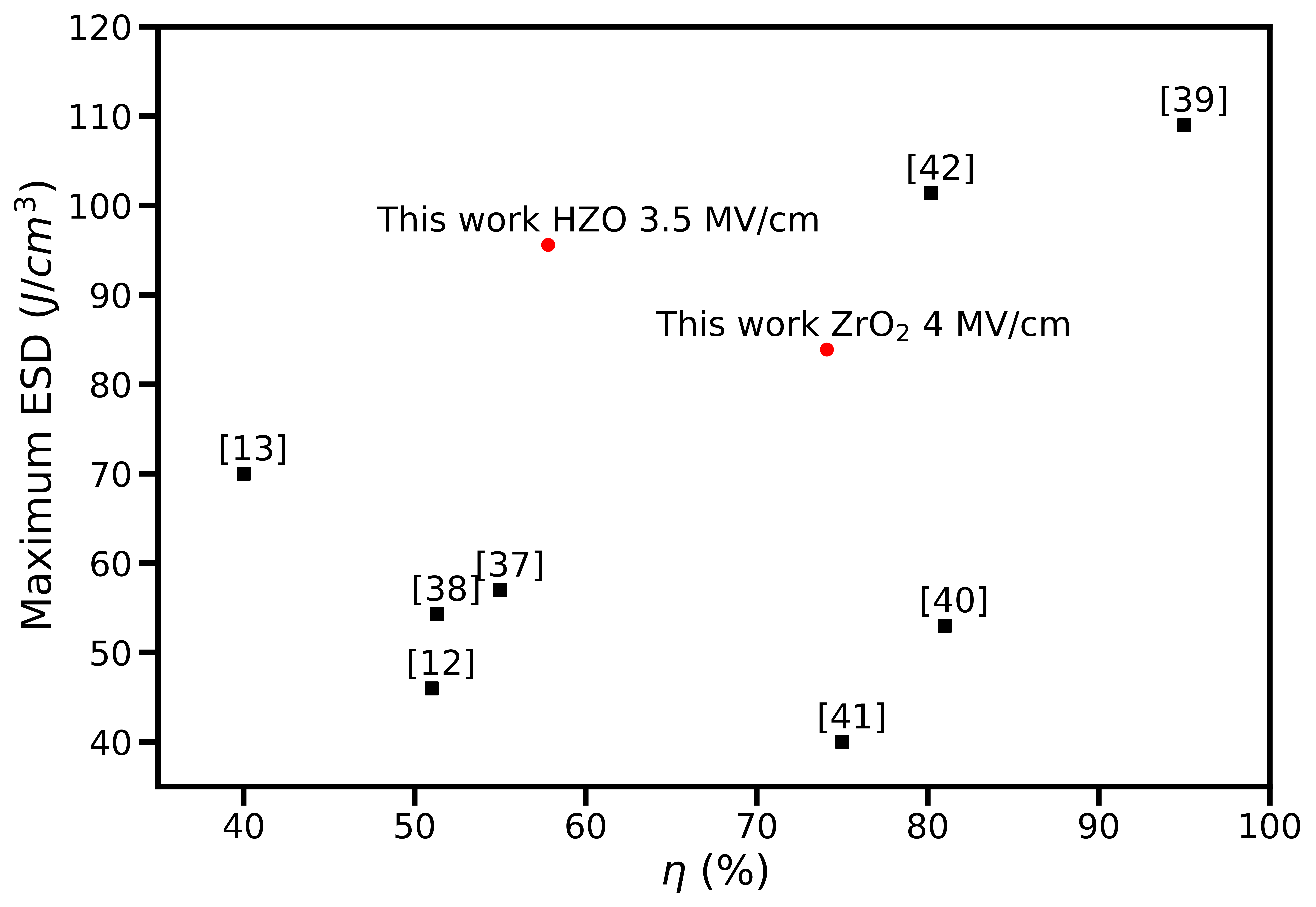}
    \caption{\label{fig:ESD_vs_Eff} Comparison of FE and AFE properties of different fluorite nanosupercapacitors from literature \cite{park2014thin,hoffmann2019negative,kozodaev2018doped,silva2020energy,do2017scale,lomenzo2017doped,kuhnel2019,SHUAI2023} and our work.}
\end{figure}

\indent
    The current literature for nanosupercapacitors using FE, AFE but also relaxor-ferroelectric (RFE) hafnium- and zirconium-based fluorite materials is compared with our results in Figure \ref{fig:ESD_vs_Eff}. One can observe that only few papers are showing higher ESD and $\eta$ than our present work. Moreover, FE \ce{HZO} is also showing excellent properties for nanosupercapacitor applications compared to what was previously observed for similar FE materials. 

\indent    
    A limiting factor to further improve the ESD and efficiency in thin films is the FE and AFE film thickness scaling. Fluorite films have limited energy storage scaling properties due to the increase of the monoclinic phase proportion at large thicknesses \cite{park2013evolution}. However, the ESD achieved for a thin film can be significantly enhanced by transitioning to a multilayered and three-dimensional (3D) structure \cite{hoffmann2019negative}, an aspect that can be explored in future studies. This transition holds the potential to elevate the ESD by several orders of magnitude, promising new avenues for enhanced energy storage capabilities. Investigations into the multilayered and 3D architectures thus represent an exciting frontier in the quest for optimizing energy storage efficiency, offering prospects for groundbreaking advancements in the field.
    
\section{Conclusion}

\indent
    We investigated the potential of ferroelectric (FE) and antiferroelectric (AFE) fluorite-structured materials, such as hafnium oxide (\ce{HfO_2}) and zirconium oxide (\ce{ZrO_2}), for energy-efficient applications, addressing the urgent need to curb the soaring energy demands of the digital age. By integrating these materials into existing CMOS technology, it demonstrates a forward-looking approach to enhancing electronic device efficiency through advanced energy conversion mechanisms. The research provides a comprehensive analysis of the structural and electrical properties of \ce{HZO} and \ce{ZrO_2} thin films, showcasing their significant potential in solid-state electrostatic energy storage. 
    
\indent
    Moreover, the study compares the energy storage performances of FE, AFE, and LD samples, underlining the superior energy storage density (ESD) and efficiency of AFE materials. This finding is critical, as it highlights the promise of AFE \ce{ZrO_2} in energy storage applications, offering a balanced trade-off between high ESD and efficiency. The meticulous methodology, from synthesis to characterization, provides a robust framework for assessing the capabilities of these materials and sets a benchmark for future studies in the field.
    
\section{Acknowledgment}

This work was carried out on the NanoLyon technology platform and implemented inside the NanOx4EStor project. We would like to specifically thank Céline Chevalier, Giovanni Alaimo-Galli and Jean-Charles Roux for their implication on the research project at the NanoLyon platform. This NanOx4EStor project has received funding under the Joint Call 2021 of the M-ERA.NET3, an ERA-NET Cofund supported by the European Union’s Horizon 2020 research and innovation program under grant agreement No 958174. This work was supported by the Portuguese Foundation for Science and Technology (FCT) in the framework of the M-ERA.NET NanOx4EStor Contract no. M-ERA-NET3/0003/2021, by Executive Agency for Higher Education, Research, Development and Innovation Funding (UEFISCDI) and by the Agence Nationale de la Recherche (ANR) under the contract ANR-22-MER3-0004-01. 
    
\section*{References}
\bibliographystyle{unsrt}
\bibliography{Biblio.bib}

\end{document}